\documentclass[twocolumn]{aastex631}

\newcommand{\GG}[1]{}

\shorttitle{Supermassive black holes and their hosts}
\shortauthors{Silverman et al.}

\begin{document}




\title{Inferences on relations between distant supermassive black holes and their hosts complemented by the galaxy fundamental plane}




\author{John D. Silverman}
\affiliation{Kavli Institute for the Physics and Mathematics of the Universe, The University of Tokyo, Kashiwa, Japan 277-8583 (Kavli IPMU, WPI)}
\affiliation{Department of Astronomy, School of Science, The University of Tokyo, 7-3-1 Hongo, Bunkyo, Tokyo 113-0033, Japan}


\author{Junyao Li}
\affiliation{Kavli Institute for the Physics and Mathematics of the Universe, The University of Tokyo, Kashiwa, Japan 277-8583 (Kavli IPMU, WPI)}
\affiliation{CAS Key Laboratory for Research in Galaxies and Cosmology, Department of Astronomy, University of Science and Technology of China, Hefei 230026, China}

\author{Xuheng Ding}
\affiliation{Kavli Institute for the Physics and Mathematics of the Universe, The University of Tokyo, Kashiwa, Japan 277-8583 (Kavli IPMU, WPI)}




\begin{abstract}

%

The realization of fundamental relations between supermassive black holes and their host galaxies would have profound implications in astrophysics. To add further context to studies of their co-evolution, an investigation is carried out to gain insight as to whether quasars and their hosts at earlier epochs follow the local relation between black hole (BH) mass and stellar velocity dispersion ($\sigma_*$). We use 584 SDSS quasars at $0.2<z<0.8$ with black hole measurements, and properties of their hosts from the Hyper Suprime-Cam Subaru Strategic Program. An inference of $\sigma_*$ is achieved for each based on the stellar mass and size of the host galaxy by using the galaxy mass fundamental plane for inactive galaxies at similar redshifts. In agreement with past studies, quasars occupy an elevated position from the local $M_{BH}-\sigma_*$ relation, considered as a flattening, while maintaining ratios of $M_{BH}/M_*$ consistent with local values. Based on a forward-modeling of the sample, we demonstrate that an evolving intrinsic $M_{BH}-\sigma_*$ relation can match the observations. However, we hypothesize that these changes may be a reflection of a non-evolving intrinsic relationship between $M_{BH}$ and $M_*$. Reassuringly, there are signs of migration onto the local $M_{BH}-\sigma_*$ for galaxies that are either massive, quiescent or compact. Thus, the majority of the bulges of quasar hosts at high redshift are in a development stage and likely to align with their black holes onto the mass scaling relation at later times.

\end{abstract}









\section{Introduction}

Supermassive black holes (SMBHs) are an enigma which surprisingly may play an important role in the evolution of galaxies \citep[e.g.,][]{Somerville2015}. Because of our inability to resolve the sphere of influence of practically all SMBHs, we look for clues from their surrounding host galaxies on the physics at work which likely instill the known relations between the mass of the black hole and the properties of its host galaxy \citep{Kormendy2013}. In particular, the closest relation observed to date is between the mass of the SMBH and the stellar velocity dispersion ($\sigma_*$) which is likely indicative of the mass and concentration of the central potential well \citep[e.g.,][]{Ferrarese2000, Gebhardt2000}. The functional form of this relation may reveal the coupling between AGN-driven outflows and the ISM of their host galaxy \citep[e.g.,][]{King2003}.  

It is important to recognize that these relations are based mostly on inactive SMBHs. Their host galaxies are no different than any other typical massive galaxies as shown by their location along the well-established fundamental plane \citep[e.g.,][]{Hopkins2007}. As a result, the stellar mass and effective radius can be used together to produce a local mass relation with similar dispersion to that based on the velocity dispersion \citep{vandenBosch2016}. This opens the question as to whether the hosts of all SMBHs, including those in an active phase and in the distant universe, follow the fundamental plane relations (and exhibit similar scaling with black hole mass).

There are many studies using AGNs and luminous quasars across cosmic time to understand how galaxies and their SMBHs migrate onto the local mass relations \citep[see][for an early effort]{Peng2006}. Besides issues with respect to measuring the black hole mass, the challenge is to disentangle the host galaxy emission from the bright glare of an AGN. This can been achieved with high signal-to-noise spectroscopy or imaging observations coupled with state-of-the-art analysis tools and consideration of observational systematic effects. Ideally, a measure of the velocity dispersion of AGN host galaxies up to high redshift is preferred given the strong correlation with black hole mass in the local universe. However, it is costly to observe $\sigma_*$ for a large sample of galaxies hosting AGN \citep[e.g.,][]{Woo2010}, particularly those at higher redshift which are faint. Even so, there have been some successful efforts \citep{Woo2006,Shen2015}. 

To alleviate the need for costly spectroscopic observing programs, the detection of the total stellar mass through image detection using wide-area and deep surveys has been successful both from space and the ground. To date, there seems to be consensus that the ratio of the black hole mass to total galaxy stellar mass is nearly constant with redshift after considering inherent selection biases \citep{Jahnke2009,Cisternas2011,Sun2015,Ding2020a}. Recently, the Hyper Suprime-Cam Subaru Strategic Program (HSC-SSP) reported on the tightest constraints on the (lack of) evolution and its coupling to the intrinsic scatter which also shows no difference with the local dispersion \citep{Li2021b}. However, there remains the question as to whether the mass of SMBHs is more tightly coupled to the total stellar mass, bulge mass, or velocity dispersion for AGN beyond the local universe. 

High-resolution imaging of quasar hosts not only provides a stellar mass measurement of the host galaxy but also an effective radius which typically is defined as the value which encompasses half the light. As implemented in this study for the first time, the stellar mass and effective radius are used in conjunction to predict the value of the velocity dispersion under the premise that quasar hosts follow the stellar mass fundamental plane of galaxies at their respective epochs; there is supporting observational evidence for the latter in the local Universe based on quasars \citep{Wolf2008} and radio AGN \citep{Bettoni2001,Woo2004,Herbert2011}. 

Coupled with the  $M_{BH}-\sigma_*$ relation, one can then construct a relation between the mass of a SMBH and the properties of its host galaxy in terms of stellar mass and effective radius as done in  \citet{vandenBosch2016} for the local universe. This means that imaging alone would be sufficient to assess the black hole mass of a particular galaxy, if this relation holds for a broad range of redshift, type of galaxy and environment. The question is whether the BH-size-stellar mass relation from \citet{vandenBosch2016} applies to higher redshift galaxies. One would think so since galaxies irrespective of type still follow the local fundamental plane \citep[e.g.,][]{Bezanson2015}. However, the local $M_{BH}-\sigma_*$ relation must hold as well.

Here, we further examine whether quasars and their host galaxies follow the local mass relations in terms of the stellar mass and effective radius. This study is possible due to the 2D decomposition of uniformly-selected quasars and their host galaxies at $0.2<z<0.8$ using the unique combination of depth, area and high spatial resolution of the HSC-SSP \citep{Li2021a}. Equally important, we construct a large simulated sample of galaxies with quasars to ensure that our results are not due to selection. The stellar mass FP studies at higher redshifts, available in the literature, are employed to gain insight into the form of the $M_{BH}-\sigma_*$ relation over the redshift range of our sample. These results then aid in the interpretation of direct observations of $\sigma_*$ from the literature. Throughout this paper we use a Hubble constant of $H_0 = 70$ km s$^{-1}$ Mpc$^{-1}$ and cosmological density parameters $\Omega_\mathrm{m} = 0.3$ and $\Omega_\Lambda = 0.7$. We assume a Chabrier initial mass function for estimates of stellar mass.

\section{Data}

\subsection{SDSS/HSC quasars}

We use 584 SDSS quasars ($0.2<z<0.8$) selected uniformly from $\sim$5000 ($z<1$; \citealt{Paris2018}) with optical imaging from the Second Data Release (PDR2) of the HSC-SSP \citep{Aihara2019}. The black hole masses are available from~\citet{Rakshit2020} and based on single-epoch spectroscopy using the virial method ~\citep{Peterson2004, Vestergaard2006}. The typical uncertainties are assessed to be around 0.4 dex. 

\citet{Li2021a} provides the stellar mass and size of their host galaxies. This is achieved by decomposing the optical images with accurate knowledge of the point spread function into the quasar and host galaxy contributions. Using {\tt lenstronomy} \citep{Birrer2018, Birrer2021}, the host galaxy is forward modeled with a single Sersic profile that returns a measure of the half-light radius, axis ratio, Sersic index, flux, and position. The robustness of the size measurements has been extensively tested using 2D simulations for both model and real galaxies \citep[see][for details]{Li2021a}. The stellar mass of host galaxy is then derived using the 5-band photometry ($grizy$) of the host galaxy (free of quasar emission) and CIGALE~\citep{Boquien2019}, a tool to fit the spectral energy distribution (SED) of astrophysical objects. Each quasar host is then classified as either star-forming or quiescent based on a rest-frame color - stellar mass diagram. We refer the reader to \citet{Li2021a,Li2021b} for full details on the construction of this sample and higher level data products

\subsection{Simulated quasar sample}
\label{sim_samples}

Following \citet{Li2021b}, our analysis requires a simulated sample of quasars and their host galaxies to understand the impact of selection on the observed relations being investigated. Here, we provide a general overview of the procedure and refer the reader to \citet[][particularly Figure 3]{Li2021b} for further details. 

Our Monte-Carlo simulation of $5\times10^5$ galaxies starts with a sampling in redshift that matches the observed distribution of the SDSS/HSC quasars. The stellar masses cover the same range as the observed sample and are distributed to match the known functional form of the stellar mass function of galaxies \citep{Muzzin2013}, separately for star-forming and quiescent cases. A large simulated sample is needed to ensure sufficient numbers of massive cases. We then randomly assign a size (i.e., half-light radius) using the best-fit relations and their intrinsic dispersion given in \citet{Li2021a} for quasar hosts. In addition, as a check, we also preform the same analysis using sizes based on Sersic fits of 1.5 million galaxies covering $\sim$100 deg$^2$ of the HSC-SSP \citep{Kawinwanichakij2021} and the same analysis tool as applied to the quasars.

We then use two relations to assign a black hole with an expected mass to each galaxy. The first assumes that the black hole mass scales linearly with the stellar mass as parameterized by \citet{Ding2020a}. Our motivation is based on the nearly constant mass ratio seen up to $z\sim1.8$ \citep{Ding2020a,Li2021b}. The second case is based on the well-established local $M_{BH}-\sigma_*$ relation. As further described below, we use the galaxy mass fundamental plane to predict $\sigma_*$ and then insert it into the function form of the   
$M_{BH}-\sigma_*$ relation given in \citep{Onken2004}. A quasar bolometric luminosity is then assigned by assuming that the sample follows the Eddington Ratio Distribution of broad-line AGN given in \citet{Schulze2015} as a Schechter function. The sample is then restricted to match the magnitude limits of the observation by determining  an $i$-band magnitude for the simulated quasars by implementing a bolometric correction of 12 \citep{Richards2006} and a $k$-correction based on a powerlaw quasar SED with spectral index $\alpha_{\nu}=-0.44$ \citep{vandenBerk2001}. In the end, we have a large sample of simulated galaxies covering the same range of the observational parameter space as the SDSS/HSC quasar sample.

\begin{figure}
\epsscale{1.25}
\plotone{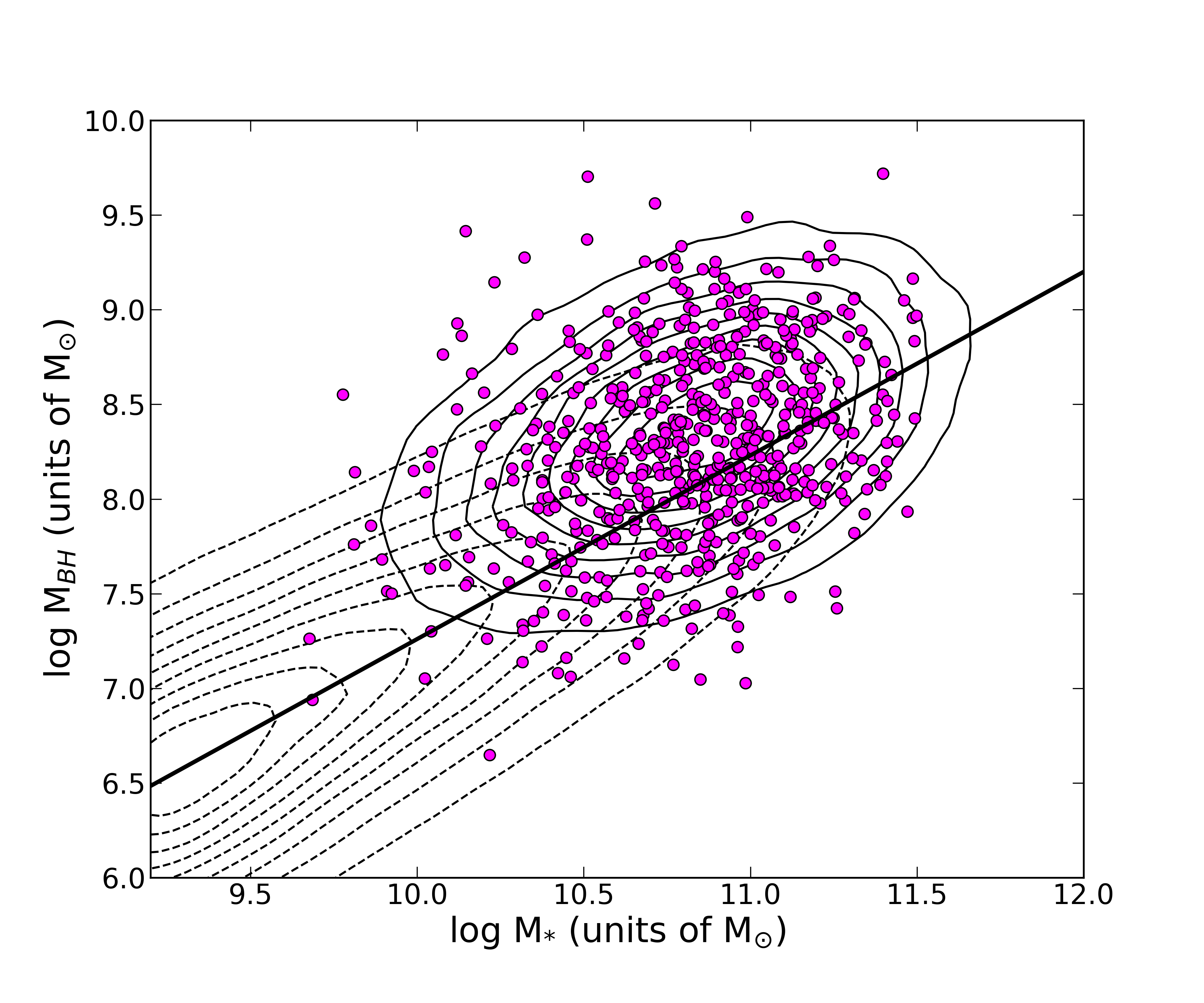}
\caption{Black hole mass as a function of stellar mass (reproduced from \citealt{Li2021b}) for the SDSS/HSC quasar sample at $0.2<z<0.8$. The local relation is shown  by the slanted black line as parameterized in \citet{Ding2020a}. Contours indicate the location of the simulated sample (Sec.~\ref{sim_samples} and ~\ref{text:selection}) with (solid) and without (dashed) selection being applied (see text for details).}
\label{fig:intro}
\end{figure}

Here, we demonstrate the impact of selection on the location of quasars in the parameter space of the $M_{BH}-M_*$ plane (Figure~\ref{fig:intro}) by showing the simulated sample before (dashed contours) and after (solid contours) observational selection has been applied. The latter is primarily the result of an imposed limit on the magnitude of the quasar which depends on which catalog it represents ($ugri$, BOSS core, eBOSS core). As reported in \citet{Li2021b}, the SDSS/HSC quasars (magenta data points) closely follow the local mass relation shown by the slanted black line. The slight offsets to higher black hole masses of the SDSS/HSC quasars are attributed to selection as detailed in the aforementioned work; this is seen by the close agreement of the parameter space spanned by the observed quasar sample and the simulated sample after selection is applied (solid contours).

\section{Initial motivation: the local relation}
\label{sec:motiv}
\begin{figure}
\epsscale{1.3}
\plotone{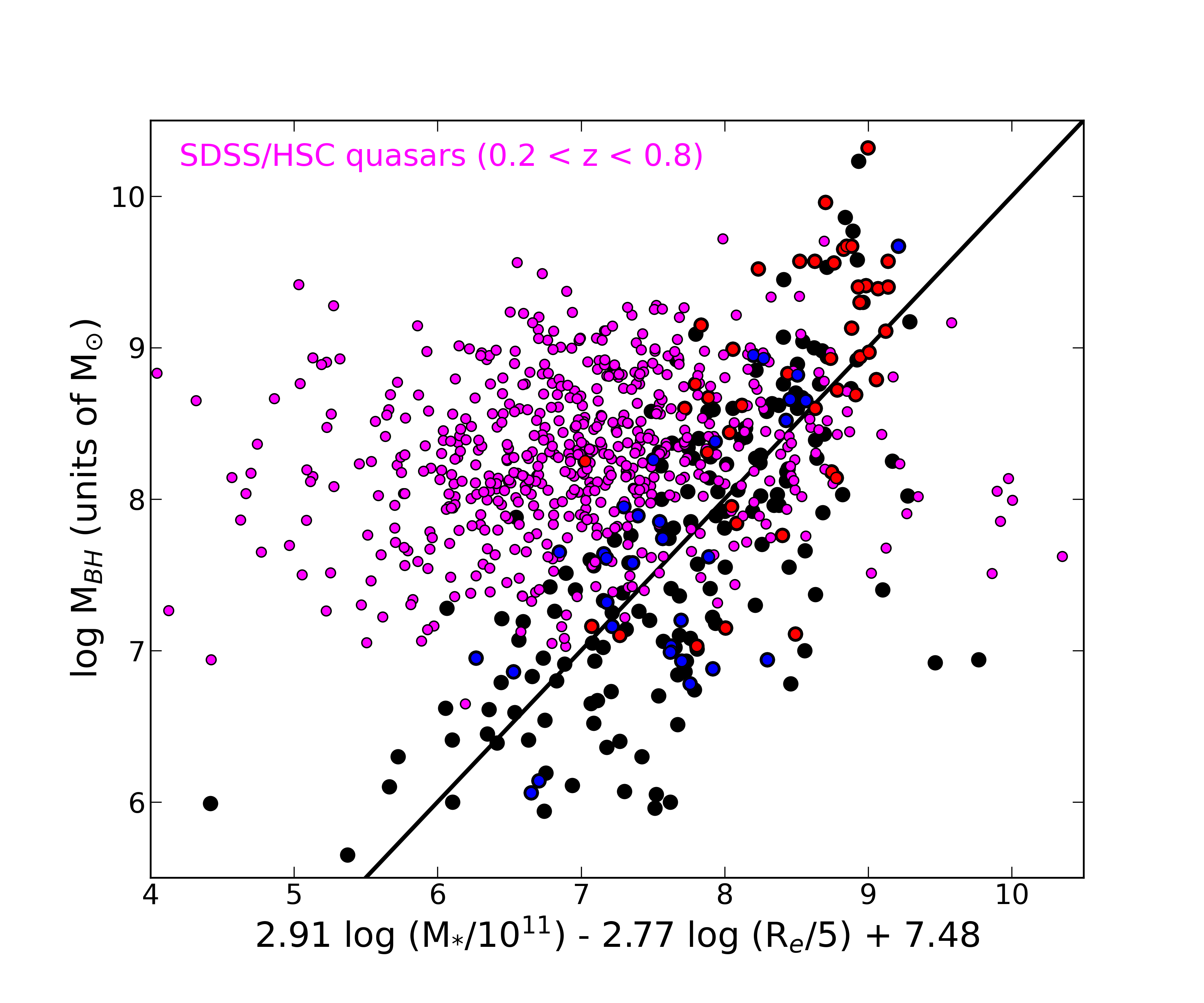}
\caption{Local scaling relation as given in \citet{vandenBosch2016} between black hole mass and a parameterization of the stellar mass and effective radius that provides an expected mass of the black hole based on local samples. The latter are color coded based on their classification, if available, in \citet{Kormendy2013} with ellipticals in red and spirals in blue. Our SDSS/HSC quasar sample is shown by the magenta circles which exhibit clear offsets compared to the best-fit local relation (slanted black line), possibly indicating that quasar hosts at higher redshift may depart from the local relations. However, an interpretation requires an assessment of selection effects and measurement uncertainties (see latter sections).}
\label{fig:fund1}
\end{figure}

We start with the formulation from \citet{vandenBosch2016} that essentially integrates two well-established relations in the local universe, namely the mass fundamental plane of galaxies and the $M_{BH}-\sigma_*$ relation, to provide a single equation between SMBH mass, stellar mass and effective radius of their host galaxy. This relation, given in Equation 5 of their study, is based on a linear fit to the local sample of galaxies. 
If quasars and their hosts at higher redshifts follow the local relations (stellar mass fundamental plane and $M_{BH}-\sigma_*$), the  \citet{vandenBosch2016} relation would be universal thus enable an independent assessment of the black hole mass for galaxies at any redshift with knowledge of their stellar mass and size. We note that \citet{vandenBosch2016} use the major axis of the isophote for the size that contains half of the light thus equivalent to the sizes from our Sersic fits.

Here, we test whether the  \citet{vandenBosch2016} relation is applicable for SMBHs at higher redshifts using our SDSS/HSC quasars that have measurements of black hole mass (from single-epoch spectra), stellar masses and effective radii with the latter two based on the 2D decomposition of HSC imaging \citep{Li2021a}. In Figure~\ref{fig:fund1}, we plot the virial (observed) black hole mass as a function of the expected black hole mass from the \citet{vandenBosch2016} relation. The SDSS/HSC quasars are shown by magenta squares. For reference, the slanted black line is the one-to-one relation and the other data points represent the galaxies used to establish the local relations. We color those from the local sample with classification as being either ellipticals or classical bulges in red, and the spirals in blue as given in \citet{Kormendy2013}.

Obviously, the SDSS/HSC quasars are significantly offset with black hole masses higher than the expected values based on the local fundamental relations. The offset is progressively larger with decreasing expected black hole mass (abscissa). The discrepancies between the high redshift quasars and the local inactive samples likely indicate that the host galaxies of quasars are either not close to being virialized systems, the local $M_{BH}-\sigma_*$ relation is not universal (i.e., fixed at all redshifts), or some combination of the two. Before we proceed in exploring the consequences of this possible breakdown in our understanding of the relation between black holes and their host galaxies, we need to evaluate such connections at the equivalent epochs and check for systematic effects due to sample selection which are well known to exist in quasar samples.

\section{Predicting $\sigma_*$ for quasar hosts using the fundamental plane of galaxies}
\label{sec:FP}
To address relations of SMBHs and their hosts at higher redshifts, we use a proxy for $\sigma_*$ applicable to our quasar sample. The stellar mass fundamental plane (FP hereafter) of galaxies enables us to infer the likely value for  $\sigma_*$ based on measurements of stellar mass and effective radius of their host galaxy where $\Sigma_* \equiv M_* / 2\pi R_e^2$.

\begin{equation}
 log~R_e=\alpha \times~log~\sigma_* + \beta \times log ~ \Sigma_* +\gamma 
 \end{equation}
 
\noindent At comparable redshifts to our quasar sample, recent studies find little change in the tilt ($\alpha$, $\beta$) and evolution of the zero-point ($\gamma$) of the stellar mass FP \citep{Bezanson2015,Zahid2016}. We employ the fixed shape and zero-point of the FP from \citet[][$\alpha=1.629$, $\beta=-0.84$, $\gamma=4.424$]{Hyde2009} with an additional offset in the zero-point of 0.042 as given in \citet{Zahid2016} based on a sample of quiescent galaxies at $z<0.6$ from the hCOSMOS survey \citep{Damjanov2018}. This characterization of the mass FP is consistent with galaxies at $z\sim0.8$ from LEGA-C \citep{Bezanson2015,deGraaf2021}, irrespective of being quiescent or star-forming. This is a key point since the majority of the host galaxies of SDSS/HSC quasars are forming stars. Equation~\ref{eq:fund_highz} gives the parameterization of the expected value of $\sigma_*$ for quasar hosts that follow the FP. 

\begin{equation}
log~\sigma_*^{p}=0.516\times log~M_* - 0.417\times log~R_e^c - 3.153
\label{eq:fund_highz}
\end{equation}

\noindent The superscript ``p" is meant to indicate that this quantity is a "predicted" value rather than being directly measured. We employ this notation throughout. The notation of the half-light radius is given with a superscript 'c' to indicate that the size here is a circularized quantity. For our SDSS/HSC quasar sample, we convert the sizes appropriately ($R_e= R_e^c\times\sqrt(1-\epsilon)$ where epsilon is the ellipticity). The sizes have been scaled to the rest-frame at 6030~${\rm \AA}$ using the relation given in \citet{vanderWel2014} which accounts for color gradients dependent on redshift and mass.

For one of the simulated samples, the values of $\sigma_*^p$ are used to derive an expected black hole mass while assuming the local relation $M_{BH}-\sigma_*$ as parametrized in \citet{Onken2004} and provided here

\begin{equation}
log~M_{BH}=4.58\times (log~\sigma_*^p - 2.30) + 8.22
\label{eq:fund_exp}
\end{equation}

\noindent and in units of $M_{\odot}$.

\section{Results: a high-z assessment of the $M-\sigma_*$ relation}

\begin{figure}
\epsscale{1.3}
\plotone{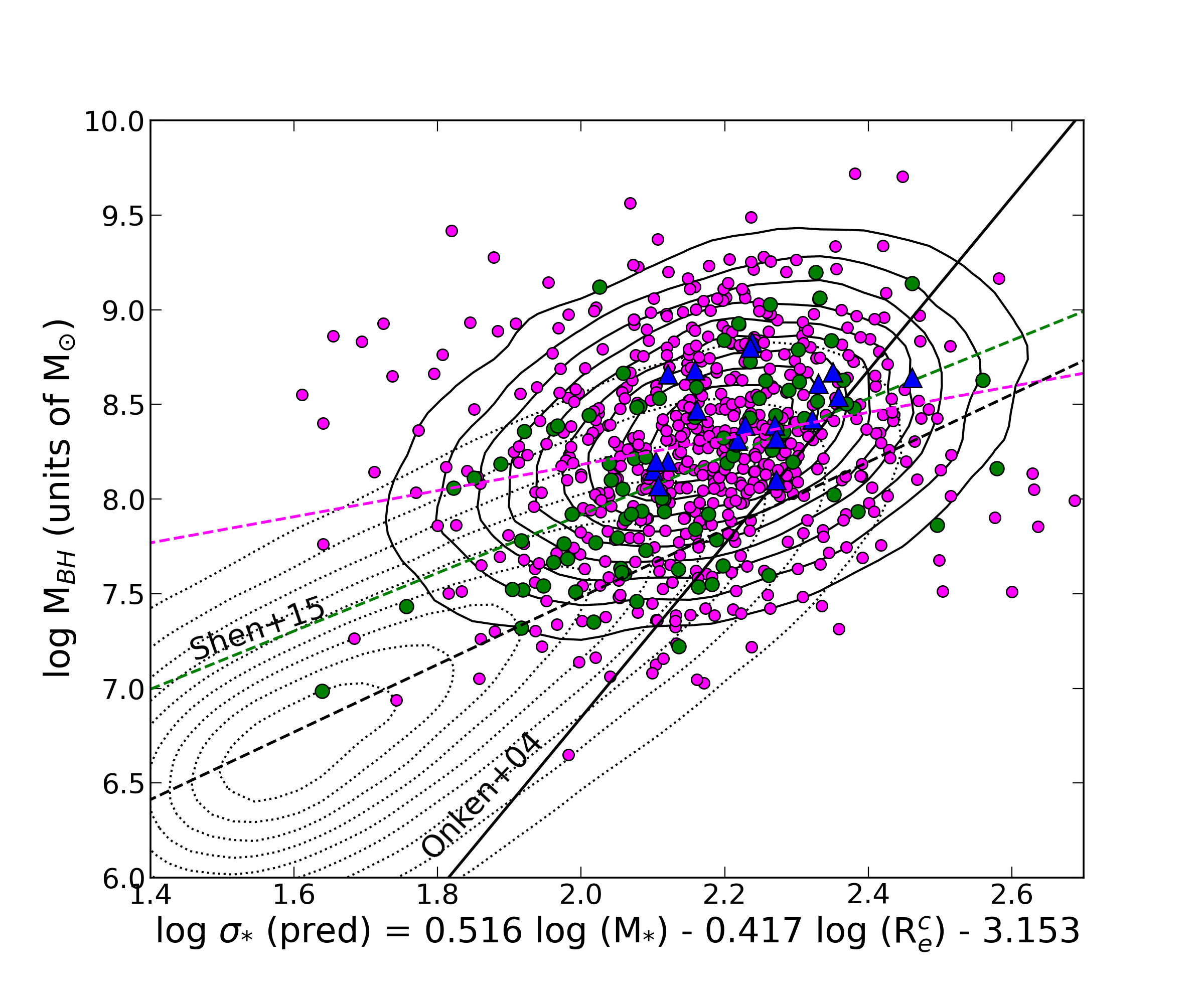}
\caption{Black hole mass ($M_{BH}$) versus the predicted stellar velocity dispersion ($\sigma_*^p$). SDSS/HSC quasars at $0.2<z<0.8$ are shown by the small magenta circles and best-fit linear relation. The simulated samples (see Sec.~\ref{text:selection}) are displayed by dashed contours for the intrinsic sample while the solid contours indicate the effect of quasar selection for the case where $log~M_{BH} \propto log~M_{*}$. The best-fit relation to the simulation data without selection applied is shown by the black dashed line. In addition, we overlay 88 quasars with direct measurements of $\sigma_*$ from the SDSS Reverberation Mapping Project (\citealt{Shen2015}; green circles and best-fit relation) and 20 Seyfert 1 galaxies (\citealt{Woo2006,Woo2008}, blue triangles).}
\label{fig:mbh-sigma}
\end{figure}

In Figure~\ref{fig:mbh-sigma}, we plot the measured black hole mass ($log~M_{BH}$) as a function of the predicted stellar velocity dispersion ($log ~\sigma_*^p$) for our quasar sample. For reference the best-fit local $M_{BH}-\sigma_*$ relation (Eq.~\ref{eq:fund_exp}) is indicated by the solid black line. Equivalent to the results presented in Section~\ref{sec:motiv}, the quasars have a distribution of black hole mass at their respective velocity dispersion that is elevated from the local relation. These offsets are larger at lower values of $\sigma_*^p$. 

To lend support to these results, we compare the location of the SDSS/HSC quasars, in the same figure, to a sample of 88 quasars at $0.1<z<1$ having direct measurements of $\sigma_*$ from high signal-to-noise spectra which are acquired through the SDSS Reverberation Mapping Project \citep{Shen2015}. In addition, \citet{Woo2006,Woo2008} provide measurements at $z\sim0.36$ and 0.56 for 20 Seyfert 1 galaxies using Keck. Reassuringly, the SDSS/HSC quasars and the samples with direct  $\sigma_*$ measurements share the same parameter space thus the offsets in the $log~M_{BH}$ - $log ~\sigma_*^p$ plane are likely accurate. 

A correlation between $M_{BH}$ and $\sigma_*^p$ is seen for the SDSS/HSC quasar sample (Fig.~\ref{fig:mbh-sigma}; magenta dashed line) which  is considerably flatter than the local relation \citep{Onken2004}. This is in general agreement with the best-fit observed relation from \citet{Shen2015}. The difference in slope is likely due to measurement uncertainties and sample selection. To further interpret these results, we use simulated quasar samples under different assumptions on the relation of black hole mass to galaxy properties (i.e., $M_*$ and $R_e$) in the next section.

\subsection{Comparison to simulated samples}
\label{text:selection}

\begin{figure}
\epsscale{1.2}
\plotone{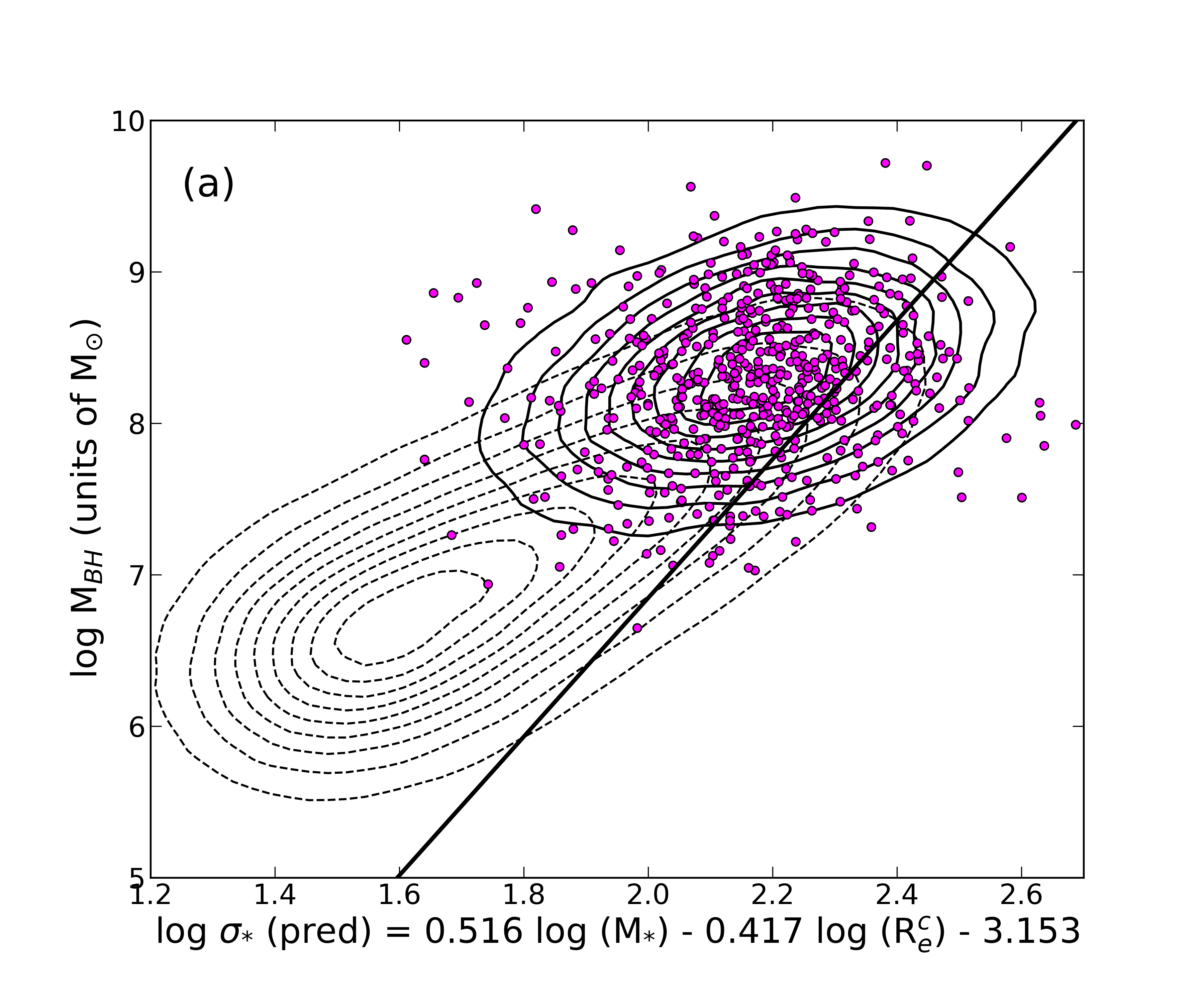}
\plotone{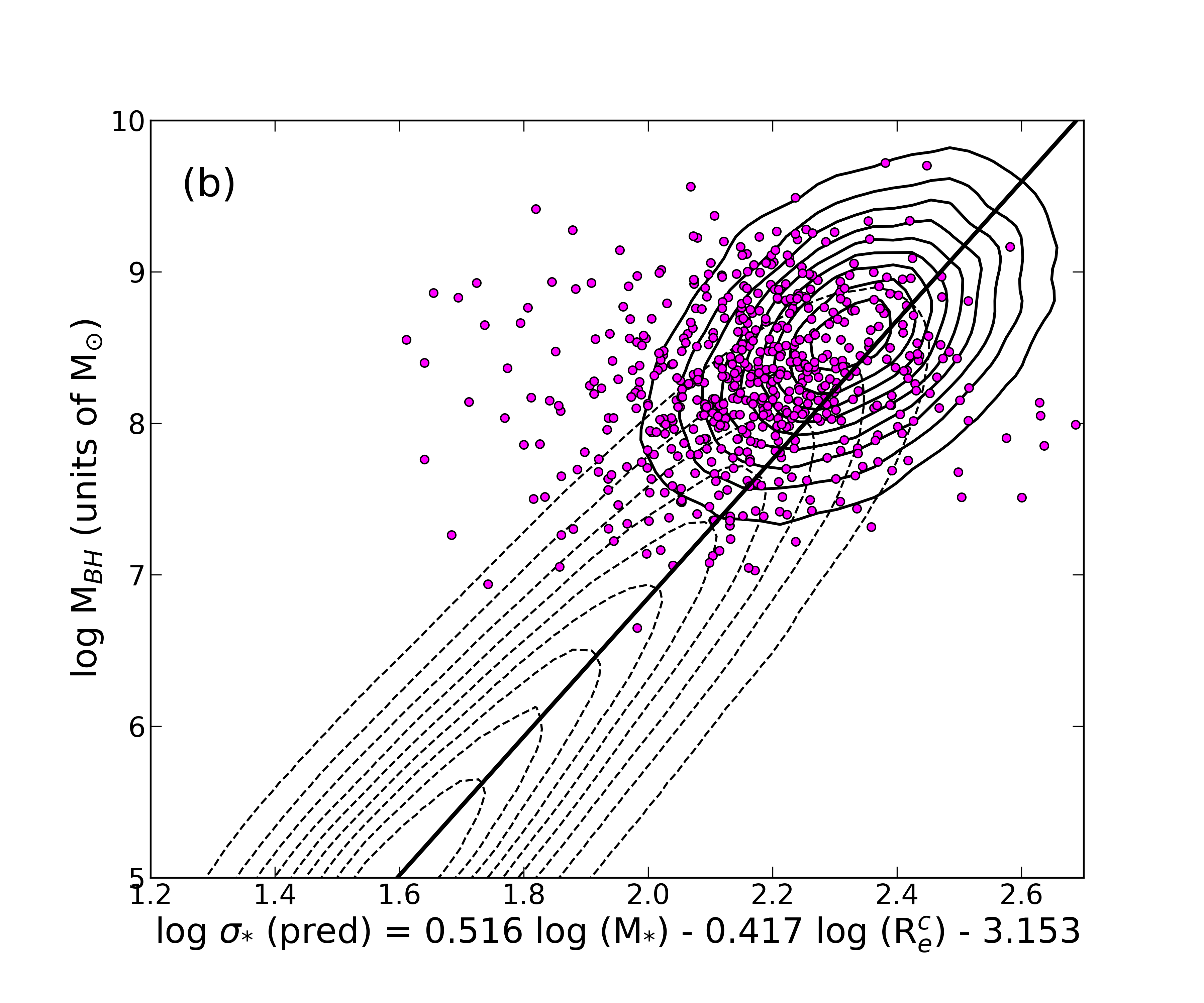}
\caption{$M_{BH}$ versus $\sigma_*^p$ as in Figure~\ref{fig:mbh-sigma}. Here, the difference between the two panels is the calculation of $M_{BH}$ (y-axis) for the simulated samples which are based on the assumption of the relation between black hole mass and host properties: (a) $log~M_{BH} \propto log~M_{*}$, and (b)  $log~M_{BH} \propto log~\sigma_*^p$.}
\label{fig:fund2}
\end{figure}

We demonstrate the impact of selection on the location of quasars in the $M_*-\sigma_*^p$ plane by showing simulated samples (as described in Sec.~\ref{sim_samples}) before and after observational selection has been applied. In Figs.~\ref{fig:mbh-sigma} and~\ref{fig:fund2}$a$, we display the simulated sample where black hole mass is assumed to be a function of only the stellar mass and follow the local $M_{BH} - M_*$ relation which has been shown to agree with our SDSS/HSC quasar sample \citep{Li2021b}. The simulated sample is shown by contours with (solid) and without (dashed) observational selection being applied. First, we find very good agreement between the location of the observed SDSS/HSC quasars, as indicated by the magenta circles, within the solid contours which depict the simulation with selection applied. Both the offsets towards lower values of $\sigma_*^p$ relative to the local relation and the dispersion of the data are similar. 

For comparison, we show in Fig.~\ref{fig:fund2}$b$ the parameter space for a different simulation where black hole mass is assumed to be a function of $\sigma_*^p$ that follows the local relation (Eq.~\ref{eq:fund_exp}). The two sets of contours reflect the application of observational selection as done in panel 'a'. In this case, there is a clear mismatch between the observed and simulated sample since the locus is off-centered with the solid contours. Therefore, the SDSS/HSC quasars and their hosts cannot follow the local relation with the observed offsets primarily due to selection effects.

\begin{figure*}
\epsscale{1.0}
\plotone{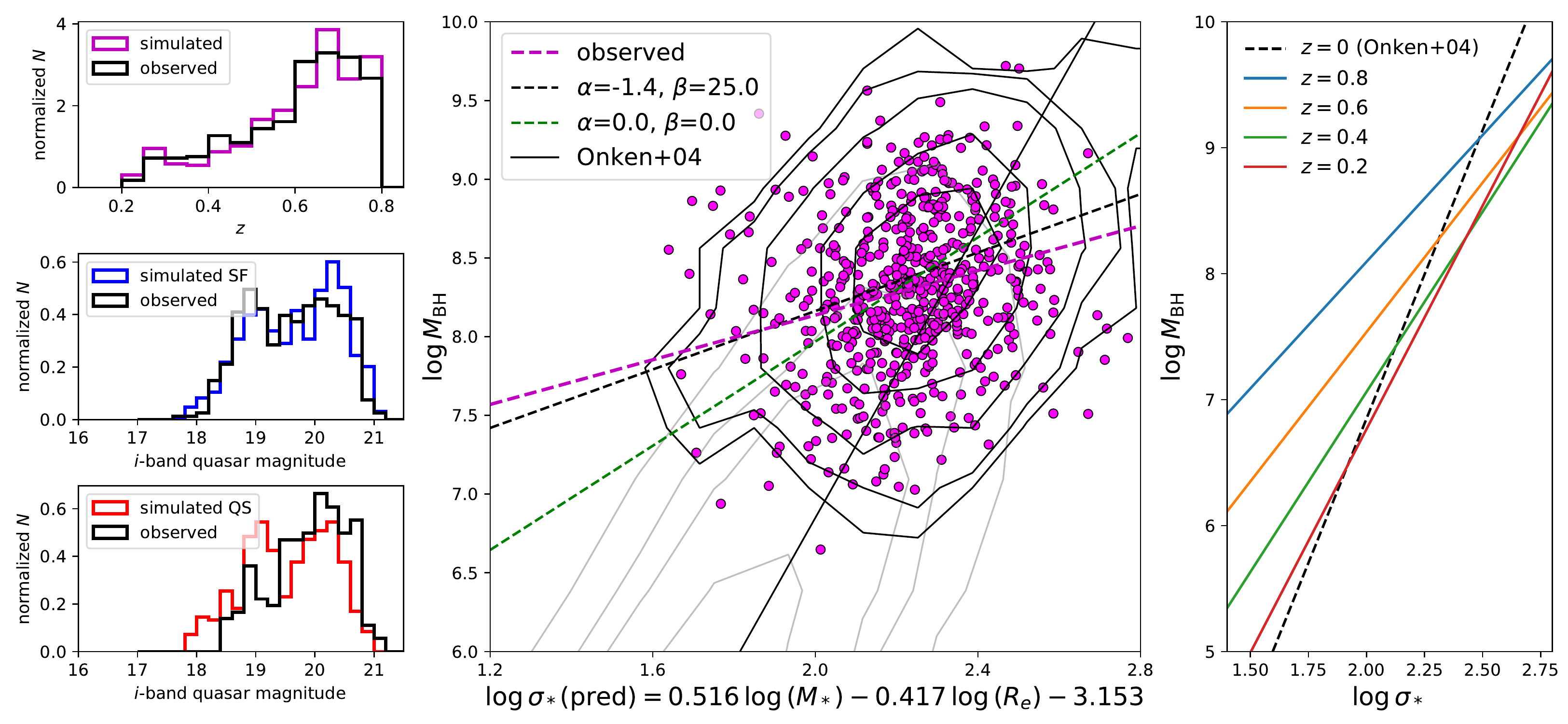}
\caption{Matching the observed (magenta data points) sample with a simulated sample based on an evolving $M_{BH}-\sigma_*$ relation. $Left$ Histograms show comparisons between the samples as a function of redshift and quasar magnitude with a separation on host galaxy type (star-forming vs. quiescent). $Middle$ $M_{BH}-\sigma_*^p$ relation. The contours show the simulated sample before (light grey) and after (black) selection is applied. The best-fit relations are shown for the SDSS/HSC quasar sample (dashed magneta line), the simulated quasar models with (dashed black line) and without (dashed green line) evolution, and the local inactive sample (solid black line). $Right$ Evolution of the $M_{BH}-\sigma_*^p$ relation in five redshift bins.}
\label{fig:msig-evol}
\end{figure*}

\subsection{An evolving $M_{BH}-\sigma_{*}$ relation}

These offsets can be interpreted as due to a flattening of the $M_{BH}-\sigma_*$ relation at higher redshift.  As shown in Fig.~\ref{fig:mbh-sigma}, the simulated sample with selection applied nicely matches the observed data when assuming a black hole to total host mass relation at high-z. As shown, the intrinsic (before observational selection is applied) simulated sample does not match the local $M_*-\sigma_*$ relation of \citet{Onken2004}. This may indicate that a flattening of the $M_{BH}-\sigma_*$ relation is inherent at high redshifts. Selection has an effect on the slope but not enough to result in the intrinsic relation matching the local $M_*-\sigma_*$ relation. This conclusion is different to that of \citet{Shen2015} who claim that selection can account for practically all of the observed flattening seen in the $M_{BH}-\sigma_{*}$ relation.

While not claiming an exact form of the intrinsic $M_{BH}-\sigma_{*}$ relation at high-z, we provide the best-fit linear relation to the simulated sample prior to any selection being applied ($log~M_{BH}=1.80\times log~\sigma_*+3.88$; black dashed line in Fig.~\ref{fig:mbh-sigma}). A measure of the intrinsic relation at high-z with accurate constraints on the slope and offset requires further analysis which carefully considers measurement uncertainties and all underlying systematic effects. For example, there are changes in the slope ($\sim$0.2) depending on the the level of uncertainty implemented for the black hole mass estimates (0.3 versus 0.4 dex) and the use of the size-mass relation for inactive galaxies from \citet{Kawinwanichakij2021}

Even so, as an exercise, we explore whether there exists a redshift-dependent model between black hole mass and $\sigma_*^p$ which can reproduce the observed parameter space of our SDSS/HSC quasar sample. We use the following relation to make a prediction on the black hole mass at a given velocity dispersion for our simulated sample.

\begin{equation}
log~M_{BH}=4.58\times(1+z)^{\alpha}\times log~\sigma_*^p -2.31 + \beta \times log(1+z)
\label{eq:fund_exp_evol}
\end{equation}

\noindent We find that $\alpha=-1.4$ and $\beta=25$ bring the simulations into broad agreement with the data (Figure~\ref{fig:msig-evol}). This model has a slope of the $M_{BH}-\sigma_*$ relation that flattens with increasing redshift and rises in normalization. Thus SMBHs are further offset above the local relation with decreasing $\sigma_*$ and redshift.

To conclude, we highlight that the offsets in the $M_{BH}-\sigma_{*}$ relation at high-z may simply be a manifestation of having a constant ratio $M_{BH}/M_*$ with redshift. As shown in Figures~\ref{fig:mbh-sigma} and ~\ref{fig:fund2}$a$, the simulated sample, prior to selection effects being applied, is in itself offset and significantly flatter than the local $M_{BH}-\sigma_{*}$ relation. Indeed, \citet{Li2021b} find that a non-evolving $M_{BH}-M_{*}$ relation bests describes the SDSS/HSC sample. This brings into question whether the black hole is more intimately connected with the velocity dispersion or total stellar mass. 

\subsection{Migration onto the local relation}

As seen in the previous figures, there is a subset of the SDSS/HSC quasars that do overlap with the local $M_*-\sigma_*$ relation. We further investigate whether there may be a trend that aligns quasars with the local relation for host galaxies of a specific property. For this exercise, we measure the difference, in log, between the observed black hole measurements and the values expected based on the predicted values of $\sigma_*^p$ (Eq.~\ref{eq:fund_highz}) and the local $M_{BH}-\sigma_*$ relation (Eq.~\ref{eq:fund_exp}). We further split the observed sample into those with hosts classified as either star-forming or quiescent as reported in \citet{Li2021a}. We note that for our purpose the offset in the mass FP of $\sim0.02$ dex between star-forming and quiescent galaxies \citep{Bezanson2015, deGraaf2021} is inconsequential thus not applied.

\begin{figure}
\epsscale{1.2}
\plotone{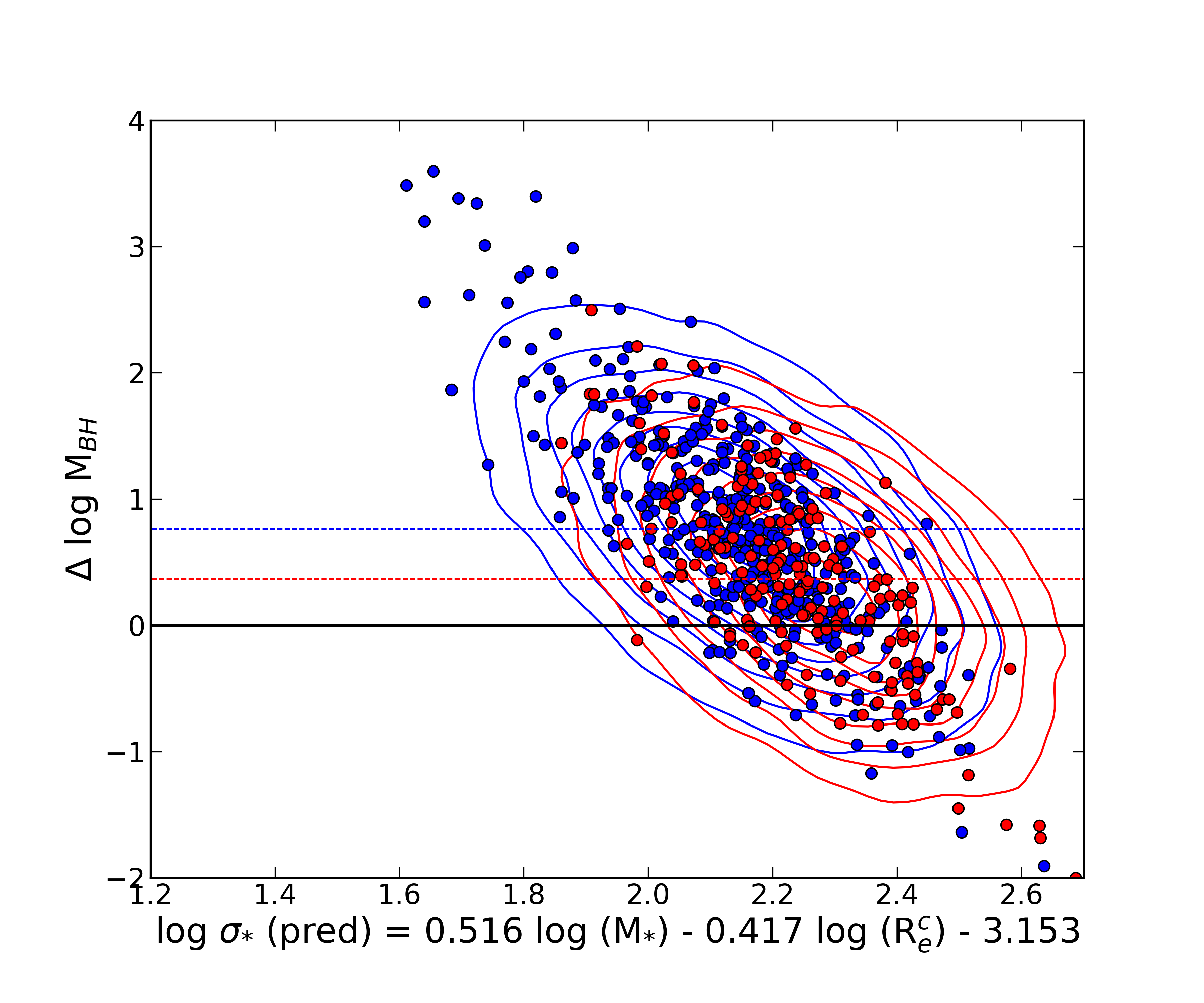}
\caption{Difference in black hole mass between the viral-based (observed) estimate and an expected value based on the local $M_*-\sigma_*$  relation. The quasar sample has been divided into those with star-forming (blue) and quiescent (red) host galaxies. The horizontal lines indicate the mean offset in black hole mass for the two samples with $log~M_*>10$ ($<\Delta M_{BH}>=0.76\pm0.01$ (star-forming) and $0.37\pm0.01$ (quiescent)). Contours show the location of the simulated samples with selection applied (blue=star-forming; red=quiescent).}\label{fig:fund4}
\end{figure}

In Figure~\ref{fig:fund4}, we plot this difference in black hole mass ($\Delta log~M_{BH}$) as a function of $\sigma_*^p$. Considering the full sample, the galaxies with the higher predicted velocity dispersions have black hole masses closer to that expected based on the local scaling relations. There is a progressive departure in black hole mass with decreasing $\sigma_*^p$. Furthermore, the quasars with quiescent hosts have black hole masses in closer agreement than the star-forming galaxies as indicate by their mean mass offset as given in the caption and labeled by the colored horizontal lines. If dividing by another factor of $R_e^c$, a similar conclusion would be reached for those that have the highest mass surface density (i.e., compactness). 

\section{Concluding remarks}

There is now mounting observational evidence (after considering selection effects and measurement errors) for a non-evolving relation between the mass of SMBHs and the total stellar mass of their host galaxies \citep{Jahnke2009,Cisternas2011,Sun2015,Ding2020a,Li2021a}. It naturally follows that SMBHs at higher redshifts have, on average, higher masses relative to the bulges of their host galaxies \citep{Ding2020a}. This is due to the fact that the fraction of stellar mass in a bulge component increases with cosmic time for the overall galaxy population \citep{Bruce2014}. In accordance with this scenario, the most massive SMBHs in the local Universe reside in bulge-dominated galaxies (i.e., the ellipticals) while those in the distant Universe are primarily hosted by galaxies with prominent stellar disks \citep{Schawinski2012,Ding2020a}. 

Here, our aim is to further illustrate such evolution between SMBHs and their hosts by using the mass and size information to infer their likely stellar velocity dispersion for quasar hosts at redshifts well beyond the samples used to establish the local mass relations. The stellar velocity dispersion is the quantity which has been recognized in the local Universe as having the tightest relation to black hole mass \citep{Kormendy2013}. Therefore, it is imperative to make comparisons between black hole mass and velocity dispersion in quasars and AGNs at higher redshifts. While there are studies that directly measure the stellar velocity dispersion \citep{Woo2006,Woo2008,Shen2015}, the sample sizes do not yet reach those comparable with wide and deep imaging surveys which cover larger samples of AGNs and quasars.

We have extended our study of a well-constructed sample of 584 SDSS quasar hosts at $0.2<z<0.8$ \citep{Li2021a,Li2021b} with optical imaging from the HSC-SSP. A decomposition of the optical emission yields a measure of the host stellar mass and size (i.e., half-light radius). The stellar velocity dispersion is then inferred based on these two quantities and a premise that quasar hosts follow the galaxy mass fundamental plane. To interpret our results, we utilize a simulated sample of SMBHs and their hosts as presented in \citet{Li2021b}. Based on our analysis, we report on the following: 

\begin{itemize}

\item The distribution of $M_{BH}$ and $\sigma_*^p$ is inconsistent with the local relation since there are significant positive offsets that increase with decreasing $\sigma_*^p$. 

\item The parameter space of $M_{BH}$ and $\sigma_*^p$ covered by our sample agrees with an independent assessment based on direct measurements of $\sigma_*$ \citep{Woo2006,Woo2008,Shen2015} using smaller samples. 

\item Based on a forward-modeling of the sample, we find that the $M_{BH}$ and $\sigma_*$ relation is intrinsically flatter than the local relation in contrast to a previous study \citep{Shen2015}.

\item While an evolutionary model of the $M_{BH} - \sigma_*^p$ relation can match the observations with a significant flattening and increase in the normalization with redshift, {\it we put forward an hypothesis that these changes in the $M_{BH} - \sigma_*^p$ relation can be naturally produced by a situation where the ratio $M_{BH}/M_{*}$ is constant up to $z\sim1$.} Therefore, a relation between black hole mass and total stellar mass may be the more fundamental link between galaxies and their black holes.
 
\item We find evidence for the migration of quasar hosts onto the local relation. The most massive galaxies and those which had their star formation quenched are more closely aligned with the local $M_{BH}-\sigma_*$ relation.
\end{itemize}

Taken together, these results support a scenario where quasar hosts are in the process of building their central mass concentration (while continuing to grow their black hole; \citealt{Li2021a,Silverman2019}). With substantial stellar mass in the disk component of their hosts, a process is required to redistribute the stars that already exist in the disk to the bulge. Also, a non-negligible fraction of the bulge mass is likely built up in-situ as evident by recent ALMA detections of centrally-concentrated star formation in high-z galaxies and AGNs \citep[e.g.,][]{Puglisi2019}. A picture is emerging where the bulge needs to be further assembled for galaxies and their SMBHs to align with the local relation which likely is regulated by AGN feedback with new additional supporting evidence with the scatter in the mass relations also showing no signs of evolution \citep{Li2021b} which agrees well with simulations invoking AGN feedback (Ding et al. in preparation).

\begin{acknowledgments}
We thank Ivana Damjanov, Rachel Bezanson and Hassen Yesuf for helpful discussions. JS is supported by JSPS KAKENHI Grant Number JP18H01251 and the World Premier International Research Center Initiative (WPI), MEXT, Japan.
\end{acknowledgments}

\bibliography{jdsrefs}{}
\bibliographystyle{aasjournal}




\end{document}